\documentclass[conference]{IEEEtran}
\IEEEoverridecommandlockouts
\usepackage{cite}
\usepackage{amsmath,amssymb,amsfonts}
\usepackage{amsthm}
\usepackage{algorithmic}
\usepackage{graphicx}
\usepackage{textcomp}
\usepackage{xcolor}
\usepackage{scrextend}
\usepackage[bookmarks=false]{hyperref} % To remove bookmarks for eXpress
\def\BibTeX{{\rm B\kern-.05em{\sc i\kern-.025em b}\kern-.08em
    T\kern-.1667em\lower.7ex\hbox{E}\kern-.125emX}}
    
\newcommand\Tstrut{\rule{0pt}{2.6ex}}         % = `top' strut

\begin{document}

\title{Active learning with RESSPECT: Resource allocation for extragalactic astronomical transients}

\author{\IEEEauthorblockN{Noble Kennamer}
\IEEEauthorblockA{\textit{Department of Computer Science} \\
\textit{University of California Irvine}\\
Irvine, USA \\
nkenname@uci.edu }
\and
\IEEEauthorblockN{Emille E. O. Ishida}
\IEEEauthorblockA{\textit{Universit\'e Clermont Auvergne} \\
\textit{CNRS/IN2P3, LPC}\\
Clermont Ferrand, France \\
emille.ishida@clermont.in2p3.fr}
\and
\IEEEauthorblockN{Santiago Gonz\'alez-Gait\'an}
\IEEEauthorblockA{\textit{CENTRA-Centro de Astrof\'isica e Gravita\c{c}\~ao} \\
%\textit{Instituto Superior Tecnico} \\
\textit{Universidade de Lisboa}\\
Lisbon, Portugal \\
gongsale@gmail.com}
\and
\IEEEauthorblockN{Rafael S. de Souza}
\IEEEauthorblockA{%\textit{Key Laboratory for Research in Galaxies and Cosmology} \\
\textit{Shanghai Astronomical Observatory}\\
\textit{Chinese Academy of Sciences}\\
Shanghai, China \\
drsouza@shao.ac.cn}
\and
\IEEEauthorblockN{Alexander Ihler}
\IEEEauthorblockA{\textit{Department of Computer Science} \\
\textit{University of California Irvine}\\
Irvine, USA \\
ihler@ics.uci.edu}
\and
\IEEEauthorblockN{Kara Ponder}
\IEEEauthorblockA{\textit{Berkeley Center for Cosmological Physics} \\
\textit{University of California Berkeley}\\
Berkeley, USA \\
0000-0002-8207-3304}
\and
\IEEEauthorblockN{Ricardo Vilalta}
\IEEEauthorblockA{\textit{Department of Computer Science} \\
\textit{University of Houston}\\
Houston, USA \\
rvilalta@uh.edu}
\and
\IEEEauthorblockN{Anais M\"oller}
\IEEEauthorblockA{\textit{Universit\'e Clermont Auvergne} \\
\textit{CNRS/IN2P3, LPC}\\
Clermont Ferrand, France \\
anais.moller@clermont.in2p3.fr}
\and
\IEEEauthorblockN{David O. Jones}
\IEEEauthorblockA{\textit{NASA Einstein Fellow} \\
\textit{University of California Santa Cruz}\\
Santa Cruz, USA \\
0000-0002-6230-0151}
\and
\IEEEauthorblockN{Mi Dai}
\IEEEauthorblockA{\textit{Department of Physics and Astronomy} \\
\textit{The Johns Hopkins University}\\
Baltimore, USA \\
0000-0002-5995-9692}
\and
\IEEEauthorblockN{Alberto Krone-Martins}
\IEEEauthorblockA{\textit{Department of Informatics}\\
%\textit{Donald Bren School of Information and Computer Sciences} \\
\textit{University of California Irvine}\\
Irvine, USA \\
algol@uci.edu}
\and
\IEEEauthorblockN{Bruno Quint}
\textit{NSF’s NOIRLab}\\
\IEEEauthorblockA{$\qquad$ \textit{Gemini Observatory} $\qquad$\\
La Serena, Chile \\
bquint@gemini.edu}
\and
\IEEEauthorblockN{Sreevarsha Sreejith}
\IEEEauthorblockA{\textit{Universit\'e Clermont Auvergne} \\
\textit{CNRS/IN2P3, LPC}\\
Clermont Ferrand, France \\
sreevarsha.sreejith@clermont.in2p3.fr}
\and
\IEEEauthorblockN{Alex I. Malz}
\IEEEauthorblockA{\textit{German Centre for Cosmological Lensing} \\
\textit{Ruhr-University Bochum}\\
Bochum, Germany \\
0000-0002-8676-1622}
\and
\IEEEauthorblockN{Llu\'is Galbany}
\IEEEauthorblockA{\textit{Dep. de F\'isica Te\'orica y del Cosmos} \\
\textit{ Universidad de Granada}\\
Granada, Spain \\
0000-0002-1296-6887}
\and
\hspace*{3cm}
\IEEEauthorblockN{(The LSST Dark Energy Science Collaboration and the COIN collaboration)}
}

% Following Lines for Copyright notice
%\IEEEoverridecommandlockouts
%\IEEEpubid{\makebox[\columnwidth]{978-1-7281-2547-3/20/\$31.00~\copyright2020 IEEE \hfill} %\hspace{\columnsep}\makebox[\columnwidth]{ }}

%\IEEEoverridecommandlockoutsit
\IEEEpubid{\makebox[\columnwidth]
{978-1-7281-2547-3/20/\$31.00~\copyright2020 IEEE \hfill}
\hspace{\columnsep}\makebox[\columnwidth]{ }}

\maketitle

%For copyright.
\IEEEpubidadjcol

\begin{abstract}
The recent increase in volume and complexity of available astronomical data has led to a wide use of supervised machine learning techniques. Active learning strategies have been proposed as an alternative to optimize the distribution of scarce labeling resources. However, due to the specific conditions in which labels can be acquired, fundamental assumptions, such as sample representativeness  and labeling cost stability cannot be fulfilled. The Recommendation System for Spectroscopic follow-up (RESSPECT) project aims to enable the construction of optimized training samples for the Rubin Observatory Legacy Survey of Space and Time (LSST), taking into account a realistic description of the astronomical data environment. In this work, we test the robustness of active learning techniques in a realistic simulated astronomical data scenario. Our experiment takes into account the evolution of training and pool samples, different costs per object, and two different sources of budget. Results show that traditional active learning strategies significantly outperform random sampling. Nevertheless, more complex batch strategies are not able to significantly overcome simple uncertainty sampling techniques. Our findings illustrate three important points: 1) active learning strategies are a powerful tool to optimize the label-acquisition task in astronomy, 2) for upcoming large surveys like LSST, such techniques allow us to tailor the construction of the  training sample for the first day of the survey, and 3) the peculiar data environment related to the detection of astronomical transients is a fertile ground that calls for the development of tailored machine learning algorithms.
\end{abstract}


%===============================================================================================
\section{Introduction}
\label{sec:intro}

Active learning techniques have been proven effective in a variety of situations where labeling is expensive or time consuming \cite{Settles12}. Nevertheless, there remains a range of real data scenarios where basic assumptions behind these techniques, such as sample representativeness and stability, are not fulfilled -- yet, the task of optimizing the allocation of limited labeling resources continues to be of paramount importance. In this work, we explore one specific scenario: the classification of extragalactic astronomical transients. 

In the last couple of decades, technological developments have led to a dramatic increase in the volume and complexity  of astronomical data. This scenario will soon escalate with the arrival of the Rubin Observatory Legacy Survey of Space and Time\footnote{\url{https://www.lsst.org/}} (LSST). LSST will produce   measurements of flux (brightness) within  broad regions of the electromagnetic spectrum (filters).  These \textit{photometric} observations can be obtained roughly in a few minutes for all sources within the telescope field of view, in effect providing a snapshot of that region of the sky at that moment in time. The survey is expected to cover the entire southern sky every few days for a total period of ten years. Nevertheless, to obtain reliable classifications, it is necessary to scrutinize each object with high resolution \textit{spectroscopic} observations. These allow the astronomer to identify the presence of individual chemical elements, which facilitates assigning it to the correct group within the astronomical zoo. This labeling process requires more telescope time (on the order of hours), a different type of instrument, and sometimes significant effort from an experienced observational astronomer who can reduce the data and translate it into a label. Although the availability of spectroscopic resources is also expected to increase during the next decade, it will always be orders of magnitude lower than its photometric counter part.

In preparation for such data deluge, the astronomical community has been investigating the application of supervised learning techniques as a strategy to provide automatic labels for thousands of objects which may never be targeted with spectroscopy \cite{Ivezic2014}. Whenever based on real data, such efforts use the available spectroscopically confirmed objects for training/validation and the final learning model is used to provide labels to the larger purely photometric sample. Despite the popularity of this approach, the intrinsically different nature of these two methods of observation results in two very different data distributions. Spectroscopy demands higher signal to noise ratio  and can only target brighter (and in many cases closer) sources. Since objects farther away exist in earlier epochs of the evolution of the universe, spectroscopic samples are restricted to certain populations of astronomical sources. In supervised machine learning applications, this mismatch translates into highly biased results \cite{Beck2017}. Moreover, the use of traditional supervised learning techniques assumes the availability of an initial training (spectroscopically confirmed) sample. This can be built from old legacy data or constructed during the first years of the survey following astronomically driven target selection strategies. In both cases, the resulting training sample will hold the biases aforementioned and consequently it will not be ideal for supervised machine learning \cite{Ishida2019}. 

 Although impossible to be completely eliminated, the discrepancy between spectroscopic (training) and photometric (target) samples can be mitigated with the help of active learning strategies \cite{solorio2005, richards2012b, Ishida2019}. Additionally, we would like to tailor the distribution of labeling resources, and consequently the construction of the training sample,  from the start of the survey. Thus ensuring that each new spectrum will add valuable information to the learning model, and not be spent on overcoming biases introduced by the initial training set. 
 
The case for classification of astronomical transients (sources which are visible  for a limited time) is even more complex. The variability of sources translates into evolving samples and labeling costs, forming a situation which is rarely addressed in the active learning literature. In  preparation for the arrival of LSST data, the LSST Dark Energy Science Collaboration\footnote{\url{https://lsstdesc.org/}} (DESC) and the Cosmostatistics Initiative\footnote{\url{https://cosmostatistics-initiative.org/}} (COIN) joined efforts in the construction of a Recommendation System for Spectroscopic follow-up (RESSPECT) -- whose goal is to guide the construction of optimized training samples for machine learning applications. This is the first public report from the RESSPECT team.

 In what follows, we focus on the problem of supernova classification and present a stress test for active learning strategies under rather realistic observational conditions. In Section \ref{sec:transients}, we describe the astronomical case in question and the data set used in our experiments. Details on how we deal with varying labeling costs and multiple sources of budgets are given in Section \ref{sec:method}. Finally, results are shown in Section \ref{sec:results} and discussed in Section \ref{sec:conclusions}. The code and data used in this work are publicly available in the COINtoolbox\footnote{\url{https://github.com/COINtoolbox/RESSPECT/tree/master}}. 

%===============================================================================================
\section{Supernova Photometric Classification}
\label{sec:transients}

In this study we focus on the classification of astronomical transients. We consider transients as stationary sources where brightness evolves with time. More specifically we are interested in supernovae, which correspond to the final stage of development of different types of stars. These are cataclysmic events, which are detected as new sources in the sky who become very bright and  remain visible for a short period (weeks to months). 

Supernovae are the origin of heavy elements in the Universe, thus playing a central role in the late cosmic evolution. Beyond their astrophysical importance, Supernovae Ia (SNe Ia) enabled the discovery of the current accelerated cosmic expansion \cite{riess1998,perlmutter1999} and remain crucial for cosmological studies. They can be recognized through unique spectral features, like the absence of hydrogen and the presence of intermediate mass elements \cite{Phillips1987}. They occur with roughly the same energy throughout the history of the Universe and thus can be used as standard candles for measuring distances at cosmological scales. 

\begin{figure}
    \centering
    \includegraphics[width=0.475\textwidth]{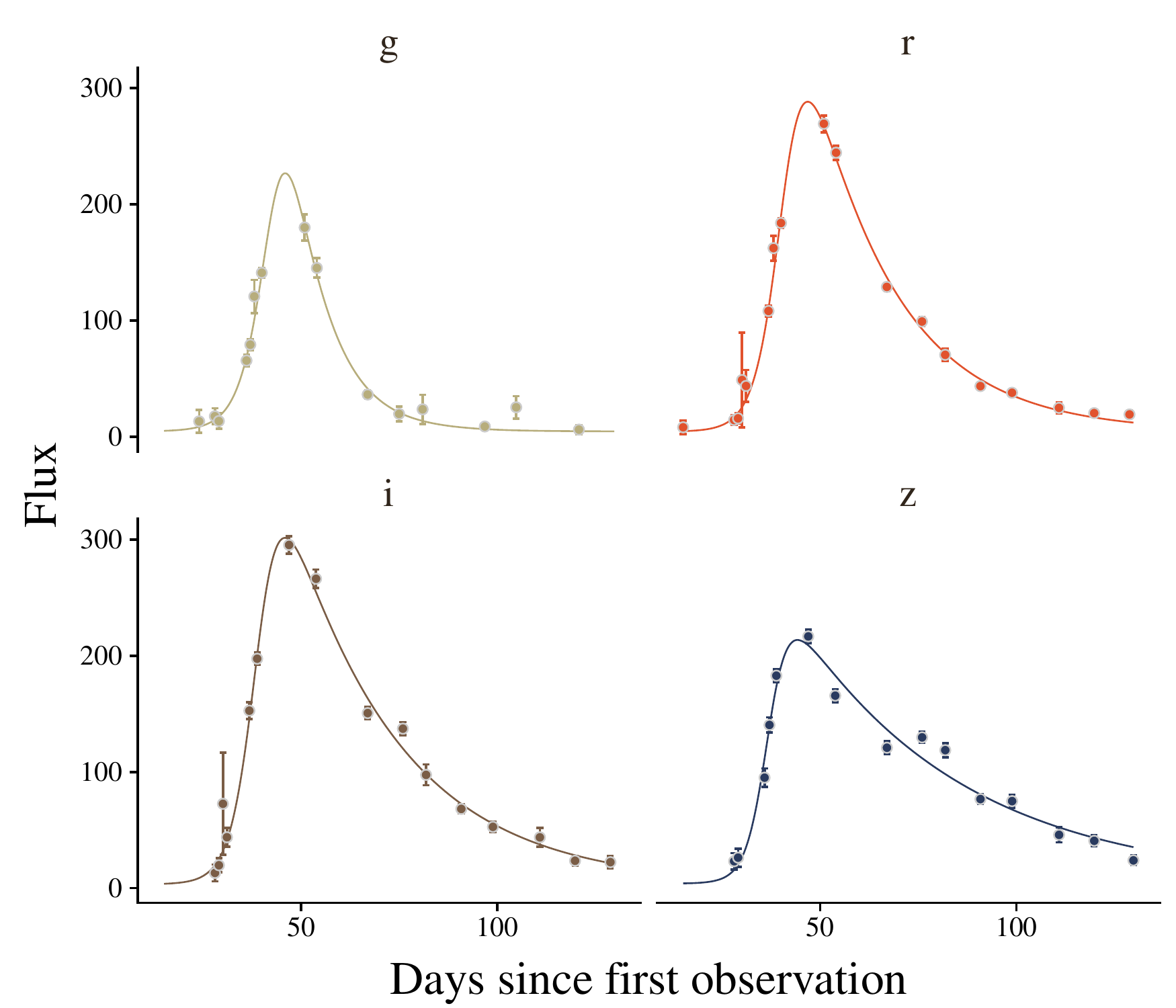}
    \caption{Light curve for a simulated type Ia supernova observed in 4 DES\footref{fn:DES} filters ([$g, r, i, z$]). The plot shows simulated flux values (points and error bars) as well as best-fit results (solid lines) across time.}
    \label{fig:lc}
\end{figure}

Since their application to cosmology became evident, SNe Ia have been among the main science drivers for many of the new generation of large scale astronomical surveys -- among them the LSST. These surveys will gather photometric measurements of the brightness evolution  (light curves) of thousands of supernovae candidates in a series of broad-band filters (see an example of simulated SN Ia light curve in Figure \ref{fig:lc}). However, since spectroscopic confirmation (labels) will only be possible for a small subset of such candidates, the potential to gain new cosmological insights from SNe Ia will rely on our ability to develop accurate automatic classifiers. 

\subsection{Caveats}
\label{subsec:caveats}

The task of classifying astronomical transients poses extra challenges beyond those faced by gathering different types of observations. We describe below some relevant issues that were considered in our experiments. Although this is not an exhaustive list, it is, to our knowledge,  a more realistic description than any other found in the literature to date.

\vspace*{2mm}
\subsubsection{Labeling window of opportunity}
\label{subsubsec:opportunity}

Once  a new source is identified as a supernova candidate, we expect its brightness to evolve and, eventually, fade away. Any spectroscopic analysis should ideally be performed when the transient is near its maximum brightness; this commonly leads to a more reliable, and less time consuming, spectroscopic confirmation. Moreover, distant or intrinsically faint targets may only be bright enough to allow spectroscopic measurements close to maximum brightness, which imposes a small time window during which labeling is possible (typically a few days). Additionally, the decision of labeling one particular target needs to be made with partial information - when one has seen only a few points in the light curve. 

\vspace*{2mm}
\subsubsection{Evolving samples}
\label{subsubsec:evolving}

In adapting the supernova classification problem to a traditional machine learning task, we build the initial training and validation/test samples using full-light curves. Our goal is to use active learning to construct a model that performs well when classifying the full light curve test sample. However, the pool sample unavoidably contains partial light curves (Section \ref{subsubsec:opportunity}). Considering, for the moment, a simplified case of fixed batches containing only 1 object: at each iteration an object is queried and sent for spectroscopic observation. Assuming the labeling process was successful, the chosen object is likely to be close to its maximum brightness phase. As a consequence, its light curve has only been partially observed. This partial light curve and its corresponding label are transferred from the pool to the training sample, which is now formed by a number of full light curve objects and one additional partial light curve. Since we expect the following day to bring some additional photometric measurements (points in the light curve) for a subset of the objects in the initial pool sample, the result is a continuous update and evolution of the training and pool samples during the entire duration of the survey.

\vspace*{2mm}
\subsubsection{Sources of budget}
\label{subsubsec:budget}

In our case study, the labeling process is extremely expensive and requires coordination between different  telescopes. The power of astronomical telescopes is proportional to the area of their primary mirror. A larger primary mirror means the telescope is able to target fainter, and consequently more distant, sources. We consider the scenario where two spectroscopic telescopes are used for labeling purposes: one telescope with a primary mirror of 4m in diameter and another with 8m. At each night, we considered 6 hours of available observation time per telescope\footnote{This is an optimistic estimation of the nightly budget.} (budget). Since spectroscopic observations of the same object require a different amount of observation time for each of the telescopes, each telescope is considered a  distinct budget source.

\subsubsection{Evolving costs per object and budget source}
\label{subsubsec:cost}

For each queried object, the time necessary to take a spectrum (which in turn can be used for labeling)  depends on the characteristics of the available spectroscopic telescope and the brightness of the target object, among other factors. As an illustration, an object with a brightness that requires $t$ minutes of spectroscopic analysis using a 4m telescope is also  a viable target for the 8m -- in which case it would require only a fraction of $t$ to complete the observation. On the other hand, a fainter object which can be observed by the 8m telescope given a large enough observation time, might not be a viable target for the 4m. Moreover, as the brightness (measured flux) of each supernova evolves with time, this cost will also depend on the time the query is made.  In our case, we update the cost  
of each queried object for the two different sources of budget (telescopes) at each active learning iteration (night). The maximum allowed observation time for any given object is set to 2 hours. Our exposure time calculator is heavily based on \cite{foster2016},  developed for the High Cadence Transient Survey  (HiTS).

\subsection{Data}
\label{subsec:data}

We used simulated data from the SuperNova Photometric Classification Challenge (SNPCC) \cite{kessler2010}. This data set was constructed to mimic observations taken by the Dark Energy Survey\footnote{\label{fn:DES} \href{https://www.darkenergysurvey.org/}{https://www.darkenergysurvey.org/}} (DES) during a period of 180 days. Distances were calculated assuming a standard cosmological model ($\Omega_m=0.3$, $Omega_{\Lambda}$=0.7, $w=-1$). Observation conditions at the telescope sight were derived from historical measurements of the ESSENCE project at the Cerro Tololo Inter-American Observatory\footnote{\url{http://www.ctio.noao.edu/noao/}} (CTIO) and incorporated to astrophysical templates. All these elements were incorporated using the  SNANA\footnote{\url{https://snana.uchicago.edu/}} software \cite{snana}, where all necessary configuration files for reproduction can be found. This data set contains three big classes of supernovae: types Ia, Ibc and II. The complete set contains 21,319 light curves in four broad-band DES filters, $[g, r, i, z]$, of which 1,103 represent a population that was spectroscopicaly confirmed\footnote{Visual description of the observational characteristics of this data set and its sub-samples are given at \cite{Ishida2019}, Figures~1,~2~and 3.}. See figure \ref{fig:lc} for an example of a typical supernova Ia light curve in 4 different filters.

%following traditional labeling strategies. It also contains the expected biases, corresponding to %closer, brighter and higher quality observations when compared to the test (photometric) %sample\footnote{Visual description of the observational characteristics of this data set and its %sub-samples are given at \cite{Ishida2019}, Figures~1,~2~and 3.}.  

\subsection{Experiment design}
\label{subsec:expdesign}

We separated our data set into 3 groups: the full training sample, identified as spectroscopically confirmed by the SNPCC data set and formed by 1,103 objects (hereafter, original training); the validation and test samples formed by 1,000 objects each, taken from the 20,216 light curves tagged as purely photometric by the SNPCC data set and following its sub-population distribution; and the pool sample comprising the remaining 18,216 objects. 

Since our pre-processing step (Section \ref{subsec:preprocessing}) requires a minimum of 5 observed points in each filter to deliver meaningful best-fit parameters, a complete input data matrix is only available starting from the 20$^{\rm th}$ day of the survey. This leaves only 160 active learning cycles (days) that we can use to build an optimal training sample. 
In order to probe the impact of the biases present in current spectroscopic samples, we also considered the situation where the initial training set is formed by only 10 objects (5 SNe Ia and 5 non-Ia) randomly chosen from the original training. This experimental configuration is also a more direct test of our active learning algorithms given that we have limited data and can only simulate the process for a small number of days.

To establish a baseline for comparison of our results, we also created a randomly sampled training set which follows closely the distribution of the validation/test sample. Results obtained when using this sample to train our learning model correspond to the best possible scenario we can achieve given our data set, labeling budget and classifier combination. The entire SNPCC data was rearranged to build this set of randomly selected training, test and validation samples (each containing 1,000 objects). The remaining objects were then allocated to a pool sample. This configuration was used to provide an upper bound to the performance. 

\subsection{Pre-processing}
\label{subsec:preprocessing}

We followed the feature extraction procedure described in \cite{Ishida2019}.
All light curves with at least 5 flux observations in each filter, were fit to the parametric function suggested by \cite{bazin2009}.
\begin{equation}
f(t) = A\frac{e^{-(t-t_0)/\tau_f}}{1+e^{(t-t_0)/\tau_r}}+B,
\label{eq:bazin}
\end{equation}
\noindent
%and performed a parametric fit to the measured flux (brightness)  as a function of time. 
The fit was performed independently for each filter.  Objects with less than 5 observed points per filter or for which the parametric fit did not converge were not included in the analysis. Figure \ref{fig:lc} shows the result of the parametric fit (full lines) along side the measured flux (points)  for a well sampled SN~Ia.  Best fit parameter values for  $p_X = \{A$, $B$, $t_0$, $\tau_f$,  $\tau_r\}$ were concatenated according to the effective wavelength of its corresponding filter, $X = [g, r, i, z]$, to form one line of the input matrix per object.

\begin{figure*}
	\centering
	\includegraphics[width=0.9\textwidth]{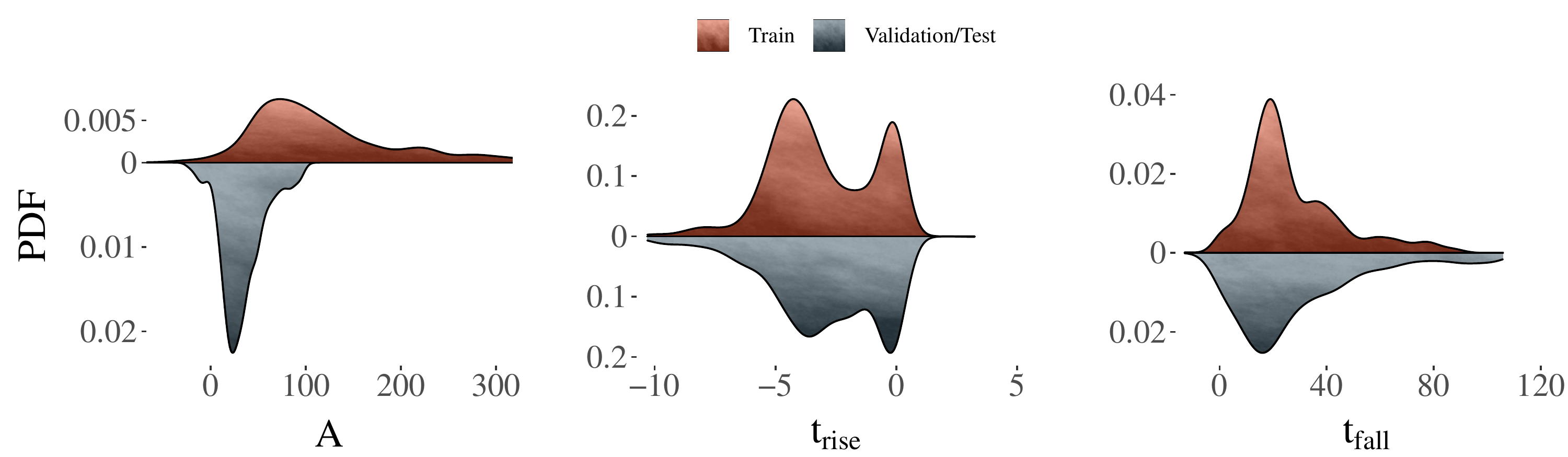}
    \caption{Comparison between light-curve features extracted from the original training (top  orange) and validation/test (bottom grey) samples in DES $r$-band.}
  \label{fig:param}
\end{figure*}

Since the initial training, validation and test samples contain full light curves, their distribution does not change. Figure \ref{fig:param} shows the distribution of best-fit parameters in $r$-band for 3 of the features considering the original training, validation and test samples. 

\begin{figure*}
	\centering
	\includegraphics[width=0.9\textwidth]{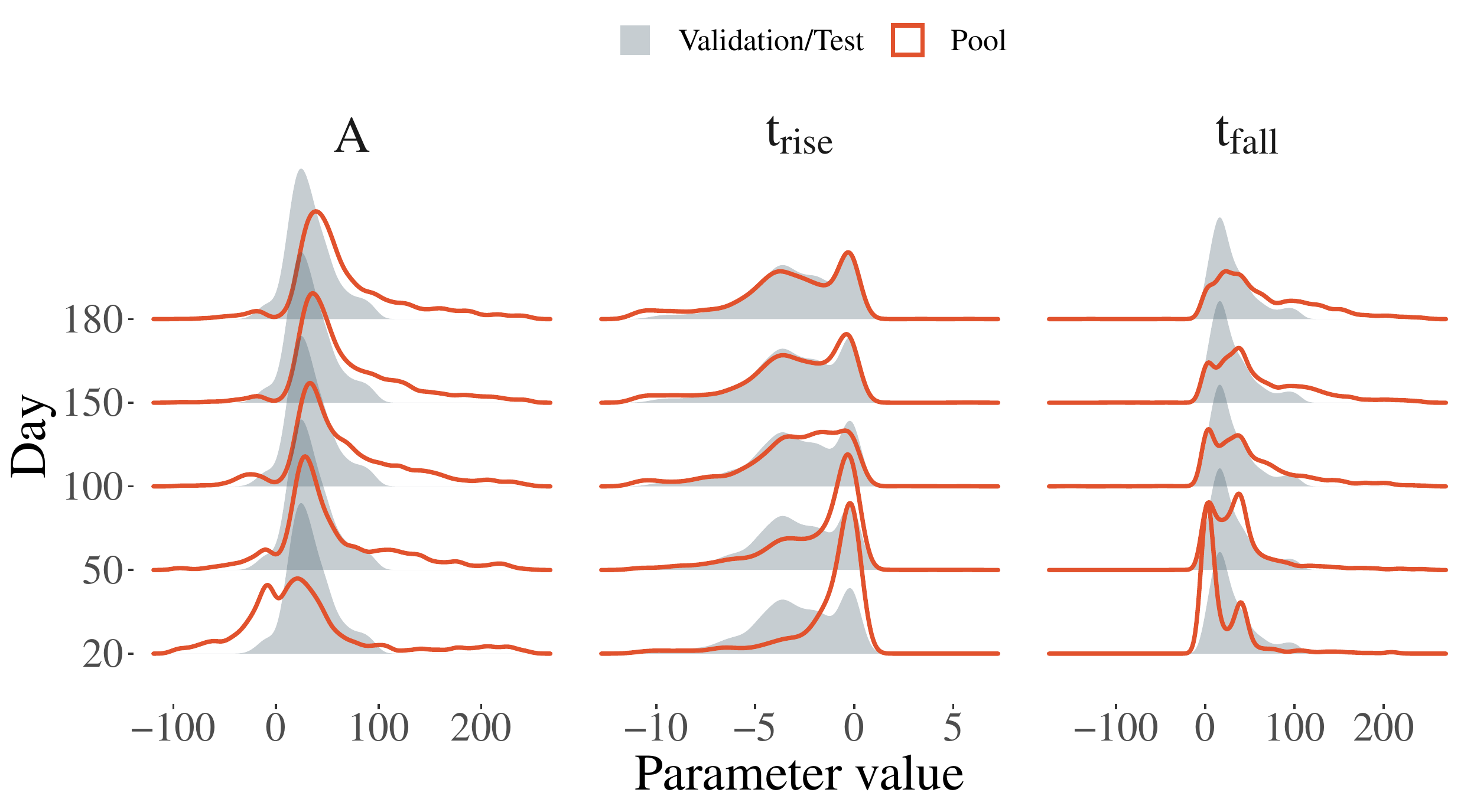}
    \caption{Distribution in feature space for the validation/test (filled grey area, fixed) and complete pool (solid orange line, evolving) samples as a function of the days since the beginning of the survey (20 to 180, from bottom to top). All distributions correspond to features extracted from $r$-band measurements.}
  \label{fig:joyQ}
\end{figure*}

For the initial pool sample the number of points observed in each light curve changes with time, thus for each day we performed the feature extraction procedure considering all light curve points observed until then. To calculate the cost of labeling, we need to estimate the brightness of the object in each day of the survey. If the last observed light curve point was measured within the last 2 days, we used that measurement as a good estimate of its current brightness. Otherwise, we use the result of the parametric fit to estimate its brightness today and use this estimate to calculate the cost of labeling with both telescopes (4m and 8m), as described in Section \ref{subsubsec:cost}. Objects bright enough to be queried by at least one of the two available telescopes form the pool sample for that day. For the 3 example features, Figure \ref{fig:joyQ} shows how the  distribution of the complete pool sample (orange) changes with the evolution of the survey in comparison with the static validation/test samples (gray) in $r$-band. 

%===============================================================================================
\section{Methodology}
\label{sec:method}

Once the training, pool, validation and test samples were properly set up (Sections~\ref{subsec:expdesign} and~\ref{subsec:preprocessing}), we recorded the performance of different active learning strategies using Random Forests~\cite{Breiman01}. For the purpose of this paper, we will only consider a binary classification problem (SN Ia/non-Ia).  For all the experiments described in Section \ref{subsec:expdesign}, we applied a naive Random Sampling (RS) strategy, where objects were randomly chosen from the pool without any selection criteria. This will serve as a lower bound for comparison with active learning techniques. 

\subsection{Active Learning Strategies}

The label constrained environment described above is a prime candidate to benefit from active learning. Moreover, it imposes significant challenges that have been under-addressed in the literature, especially from an empirical stance (Section \ref{subsec:caveats}). First, the population that can be spectroscopically observed will always differ from the target population. This requires active learning to perform well when the pool set is not representative of the validation and test sets. Second, we must choose to label a light curve before fully observing it -- since the object must be observed near maximum brightness. Finally we must also include non-constant costs in the selection of our batch sizes. Most active learning strategies assume constant costs and thus restrict the queried batch to a fixed size per iteration -- these are known as \textit{cardinality constraints}. In our case, each object has a different cost (time necessary to get a label) and our total budget is constrained by the number of hours of spectroscopic telescope time  available per night. These are known as \textit{knapsack constraints} and have been studied in the context of discrete optimization \cite{krause2008near, krause2014submodular}. These challenges make our work an excellent case study to stress test how standard and commonly used active learning algorithms hold up to real world conditions and using modern machine learning classifiers. 

We formulate our problem in terms of pool-based active learning,  coupled with  uncertainty sampling driven techniques \cite{Settles12}. Specifically we used query by committee by performing bagging over a random forest classifier \cite{freund1997selective,  seung1992statistical, breiman1996bagging}. 
Query by committee is a known active learning strategy that invokes a set of classifiers (committee) for each object's label estimation. In this context, the queried object will be the one that exhibits strong  disagreement between the members of the committee. In bagging,  the training data is sub-sampled with replacement and each subset is used to train a different model (using Random Forests) -- each of these models is then considered a member of the committee. The criteria used to quantify the disagreement between the output of committee members is called a query selection strategy. In all experiments presented here we considered a committee of size 10, each composed of 100 trees, and only varied the query selection strategy. 

Let \((x, y)\) denote feature and label pairs where in our  case \(x\) corresponds to the concatenated best fit parameters (Equation \eqref{eq:bazin}) for the 4 DES filters, measured from a single object and \(y\) is a binary label identifying Ia/non-Ia SNe. Let \(P_\theta (y | x) \) denote the predictive probability output from a single committee member, where $\theta$ encompasses the parameters of the learning model. Since each member of the committee generates a predictive probability over the estimated class, we can define the average committee predictive probability as

\begin{equation}
    P_C(y | x) = \dfrac{1}{N_C} \sum_{c} P_{\theta_c} (y | x),
\end{equation}
\noindent where \(N_C\) is the committee size and the sum runs over all committee members. We use this distribution to build all other selection strategies. 

One of the most common selection strategies is the soft vote entropy \cite{Settles12}. In information theory, entropy measures the expected (average) amount of information uncovered by identifying the outcome of a random trial \cite{mackay2003information}. In this context, if a given object has a high probability of belonging to a given class, it is unlikely that  labeling it will add new information to the model. On the other hand, if an object has equal probability of belonging to all possible classes, labeling it will uncover currently missing information and certainly improve our model. Considering the prediction of each committee member as a vote, this strategy will choose to query the object with highest entropy among all committee members. Mathematically, we have

\begin{equation}
\label{eq:entropy}
 x^* = \arg\max_x \left(- \sum_{y} P_C(y|x)\log P_C(y|x)\right),
\end{equation}
\noindent where \(x^{*}\) is the queried object\footnote{For a binary classification problem, this is equivalent to the uncertainty sampling strategy used in \cite{Ishida2019}.}.

We also use the average Kullback-Leibler (KL) divergence between the individual committee members and the average committee probability as a query selection strategy \cite{kullback1951information, Settles12},
\begin{equation}
\label{eq:kl}
 x^* = \arg\max_x \left(\dfrac{1}{N_C} \sum_{c} KL(P_{\theta_c}(y|x) ||  P_C(y|x))\right).
\end{equation}

\noindent Thus, selecting the objects with the most disagreement among the committee members. This selection strategy is equivalent to the one defined by Bayesian Active Learning by Disagreement (BALD) \cite{houlsby2011bayesian}. To our knowledge the equivalence between these two strategies has not been addressed in the literature and we provide a proof in Appendix \ref{ap:BALD}.

\subsection{Batch Strategies}

The query strategies described above target one individual object per active learning cycle. When moving to batch queries (targeting multiple objects per night), these strategies can face serious challenges, such as querying redundant data points \cite{Settles12}. The problem of querying diverse batches can typically be framed as a discrete optimization problem and is known to be computationally challenging. However, in practical applications, selecting multiple queries at a time is a requirement. Here we assume constant cost of acquisition across all data points; this requirement will be relaxed in the next subsection. An efficient approach if the query selection strategy is monotonic submodular, is to use a greedy algorithm which provides batches with a \((1- 1/e)\) approximation to the optimal solution \cite{nemhauser1978analysis, krause2008near}. Both of the query strategies given above are monotonic submodular \cite{kirsch2019batchbald}. While in \cite{kirsch2019batchbald} this technique was called BatchBALD, we refer to it as \textit{BatchKL} since our technique for approximating the disagreement region is not Bayesian.

Let the sets \(x_1, ..., x_b\) and \(y_1, ..., y_b\) be denoted as \(x_{1:b}\) and \(y_{1:b}\), where $b$ is the batch size. Using the definition of mutual information, $\mathcal{I}$, for two sets of random variables we have,

\begin{equation}
\begin{split}
\label{eq:BatchBald}
    \mathcal{I}(y_{1:b}, \theta | x_{1:b}, \mathcal{D}_{train}) = \ 
    &\mathcal{H}(y_{1:b}| x_{1:b}, \mathcal{D}_{train}) - \\
    &\mathbb{E}_{p(\theta | \mathbf{D}_{train})} \mathcal{H}(y_{1:b}| x_{1:b}, \theta, \mathcal{D}_{train}),
\end{split}
\end{equation}
\noindent where \(\mathcal{H}\) refers to entropy, $\mathbf{D}_{train}$ the training data and $\mathbb{E}$ is an expectation. The mutual information can be seen as the intersection of the information content between two sets of random variables \cite{yeung1991new}. This strategy accounts for overlaps in the information content between different data points, \(x_{1:b}\),  and model parameters, \(\theta\). By accounting for these overlaps we can avoid querying redundant data points. This function is monotonic submodular and thus, when optimized with a greedy algorithm, provides a \((1- 1/e)\) approximation to the optimal solution \cite{kirsch2019batchbald}. We use equation \eqref{eq:BatchBald} to define the BatchKL strategy as:

\begin{equation}
\label{eq:BatchBald_Selection}
    x_{1:b}^{*} = \arg\max_{x_{1:b}} \mathcal{I}(y_{1:b}, \theta | x_{1:b}, \mathcal{D}_{train}).
\end{equation}

Note that the first term on the right hand side of equation \eqref{eq:BatchBald}, the joint entropy, is also monotonic submodular. We use it to define the strategy we call BatchEntropy:

\begin{equation}
    x_{1:b}^{*} = \arg\max_{x_{1:b}} \mathcal{H}(y_{1:b}| x_{1:b}, \mathcal{D}_{train}).
\end{equation}

In addition to these two batch strategies we will also test a strategy that takes the top \(b\)  points from equation \eqref{eq:entropy}. We will refer to this strategy as Uncertainty Sampling Entropy (USE).

\subsection{Non-Constant Cost}
\label{subsec:nonconstcost}

As mentioned previously, each object in our pool sample has a different cost (telescope time required for labeling). In addition, our budget (telescope time) is very limited and needs to be used as efficiently as possible. We assume we have access to 6 hours of observation in 4m-class telescopes and 6 hours in 8m-class telescopes per night. The batch strategies defined in the last section assumed cardinality constraints where all objects had identical costs. We now consider the case where each object has different cost and we have a fixed budget each night (knapsack constraints \cite{krause2014submodular}). We show results where we fill up objects to each telescope, without considering their individual cost, until the budget of each telescope is full. We first assign objects to the 4m telescope until the budget is exhausted, at which point objects are assigned to the 8m telescope. We also tested strategies where we scale the query metrics by the cost of each object and greedily select objects after scaling\footnote{For more detail on these approaches see \cite{krause2008optimizing}, Chapter 5.}. However, we do not include these results as they were nearly identical to the simpler approach. 

%===============================================================================================
\section{Results}
\label{sec:results}

The performance of our results in the test sample are reported following the metrics proposed by \cite{kessler2010}, 

\begin{equation}
\textrm{accuracy (acc)} = \frac{N_{\textrm{sc}}}{N_{\textrm{tot}}}, 
\label{eq:acc}
\end{equation}
\begin{equation}
\textrm{efficiency (eff)} =  \frac{N_{\textrm{sc,Ia}}}{N_{\textrm{tot,Ia}}}, \label{eq:eff}
\end{equation}
\begin{equation}
\textrm{purity (pur)} =  \frac{N_{\textrm{sc,Ia}}}{N_{\textrm{sc,Ia}} + N_{\textrm{wc,nIa}}}, \qquad \textrm{and}  \label{eq:pur}
\end{equation}
\begin{equation}
\textrm{figure of merit (FoM)} = \frac{N_{\textrm{sc,Ia}}}{N_{\textrm{tot,Ia}}} \times \frac{N_{\textrm{sc,Ia}}}{N_{\textrm{sc,Ia}} + W N_{\textrm{wc,nIa}}}\label{eq:fom}, 
\end{equation}

\noindent where $N_{\textrm{sc}}$ is the total number of successful classifications, $N_{\textrm{tot}}$ is the total number of objects in the test sample, $N_{\textrm{sc,Ia}}$ is the number of successfully  classified SNe Ia (true positives), $N_{\textrm{tot,Ia}}$ is the total number of SNe Ia in the test sample, $N_{\textrm{wc,nIa}}$ is the number of non-Ia SNe wrongly classified as Ia (false positives) and $W=3$ is a factor that penalizes the occurrence of false positives. In our study, a false positive can have a more drastic consequence than a false negative.  In case we wrongly classify a SN Ia as non-Ia, we will lose the opportunity to use this object in our photometric cosmology analysis. However, if a non-Ia is mistakenly classified as a SN Ia, it will bias our distance estimates and, consequently, cosmological results. The figure of merit and the $W$ parameter were set to ensure that preference is given to results with high purity, without compromising efficiency. We search for the learning strategy that can maximize the figure of merit. For all the experiments described below we consider non-constant costs described in Section \ref{subsec:nonconstcost}.

\begin{table}[ht]
    \centering
    \caption{Performance metrics for the different active learning strategies when the entire SNPCC spectroscopic (training, 1103 objects) sample is given at the beginning of the survey. The table show results metric values 180 days after the start of the survey.}
    \begin{tabular}{cccccc} 
    Metric & \multicolumn{5}{c}{Learning Strategy}  \\
    \hline \Tstrut
     & RS & BatchEntropy & BatchKL & USE\\
     \cline{2-6} \Tstrut
    Accuracy        & 0.87 & 0.87 & 0.87 & 0.88\\
    Efficiency      & 0.57 & 0.59 & 0.56 & 0.66\\ 
    Purity          & 0.78 & 0.77 & 0.82 & 0.77\\
    Figure of Merit & 0.31 & 0.32 & 0.34 & \textbf{0.35}\\
    & & & & & \\
    %\hline \Tstrut
    %Size of final training  & & & & & \\
    \end{tabular}
    
    \label{tab:train_original}
\end{table}

\begin{table}[ht]
    \centering 
    \caption{Performance metrics for the different active learning strategies begining from a random initial training sample of 10 objects (5 SNe Ia, 5 non-Ias). The table shows results 180 days after the start of the survey.}
    \begin{tabular}{cccccc} 
    Metric & \multicolumn{5}{c}{Learning Strategy}  \\
    \hline \Tstrut
    & RS & BatchEntropy & BatchKL & USE\\
     \cline{2-6} \Tstrut
    Accuracy        & 0.85 & 0.87 & 0.87 & 0.87\\
    Efficiency      & 0.42 & 0.54 & 0.50 & 0.55\\ 
    Purity          & 0.85 & 0.84 & 0.83 & 0.80\\
    Figure of Merit & 0.27 & \textbf{0.34} & 0.31 & 0.32\\
    %\hline \Tstrut
    %Size of final training & 1795 & 1896 & 1868 & 1874 \\
     & & & & & \\
    \end{tabular}
    \label{tab:train10}
\end{table}

\begin{figure}
    \centering
    \includegraphics[width=0.475\textwidth]{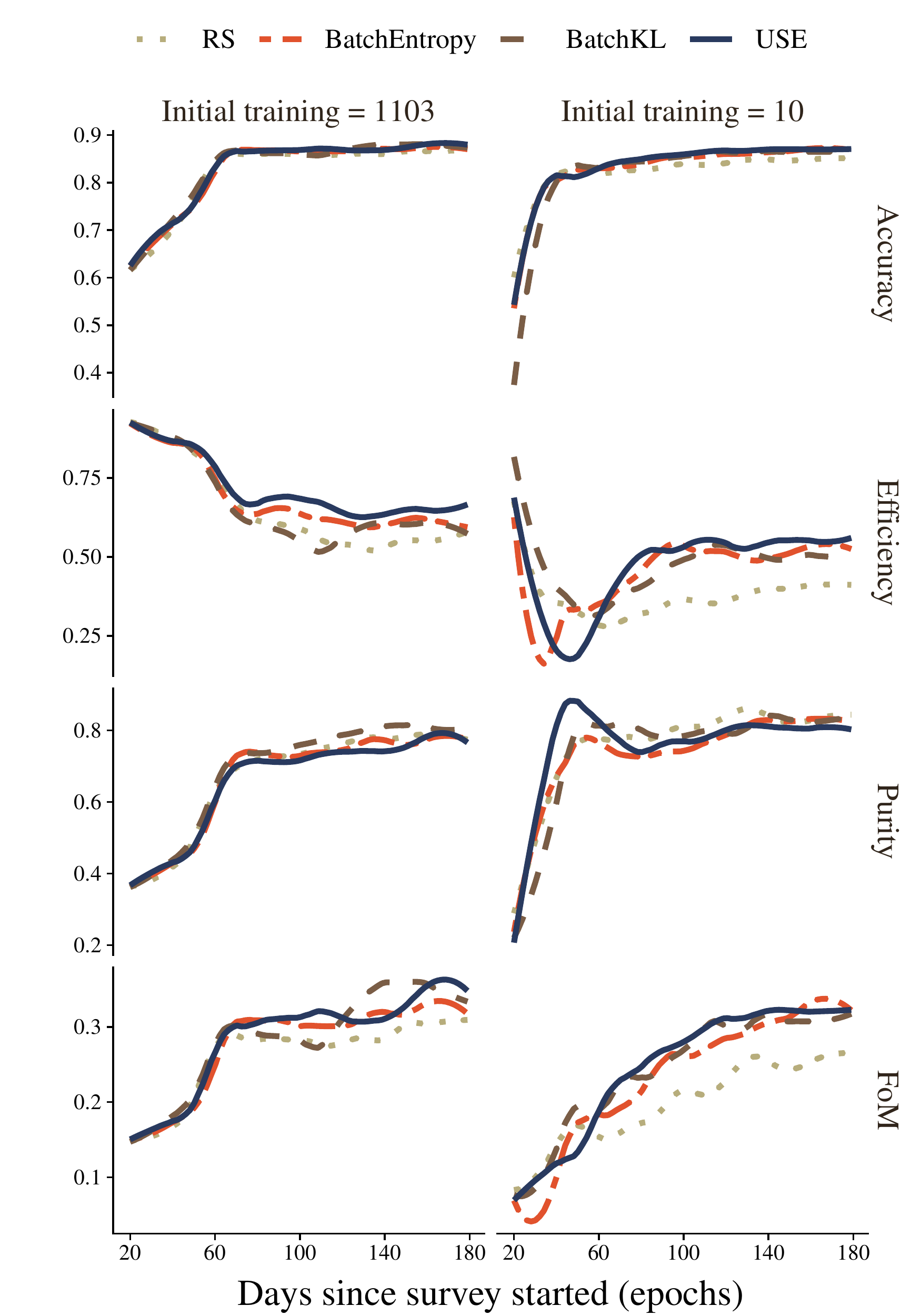}
    \caption{Evolution of different performance metrics as a function of the number of days since survey started (active learning iterations) for different learning strategies. All strategies shown here considered non-constant costs.  \textbf{Left}: initial training corresponding to the original SNPCC spectroscopic sample. \textbf{Right}: initial training containing only 10 objects (5 SNe-Ia, 5 SNe-nonIa) randomly chosen from the original SNPCC spectroscopic sample.
    }
    \label{fig:metric}
\end{figure}

% As described in section \ref{subsec:expdesign}, we present three experiments. Each one is run simulating 180 days of observation. Upon each day, our active learning strategies select a set of objects to label and add them to the training set -- while staying within the nightly budget. Experiments only differ on how the initial training set was constructed.

For our first experiment we started from the idealized case of a randomly sampled training, validation and test samples, each containing 1,000 objects. The goal of this exercise was to quantify a set of optimal results given our data, classifier and labeling resources. We used a RS strategy for the entire duration of the survey.  After 160 iterations (180 days of observation), we obtained \{acc, eff, pur, FoM\} $=$ [0.88, 0.62, 0.82, 0.37]. 

We then considered the case where the original SNPCC spectroscopic sample was completely available at the beginning of the survey, thus starting with a training sample of 1,103 objects. We applied RS, BatchEntropy, BatchKL and USE strategies and ran them through all available observation days. The behavior of the diagnostic metrics as a function of the number of active learning iterations (days since the beginning of the survey) is shown in Figure~\ref{fig:metric} (left column). Numerical values for the final state of these models are reported in Table~\ref{tab:train_original}. After 160 iterations, the final training sample had grown by $\approx 1800$ objects (for a total of $\approx 2900$). Observing the behavior of different strategies in Figure \ref{fig:metric} (left column), we see an improvement in all metrics. However, the difference in FoM results between RS and the best performing active learning strategy (USE) is merely $\approx 13\%$ ($0.04$); active learning strategies struggle to outperform RS. 

\begin{figure*}
	\centering
	\includegraphics[width=0.9\textwidth]{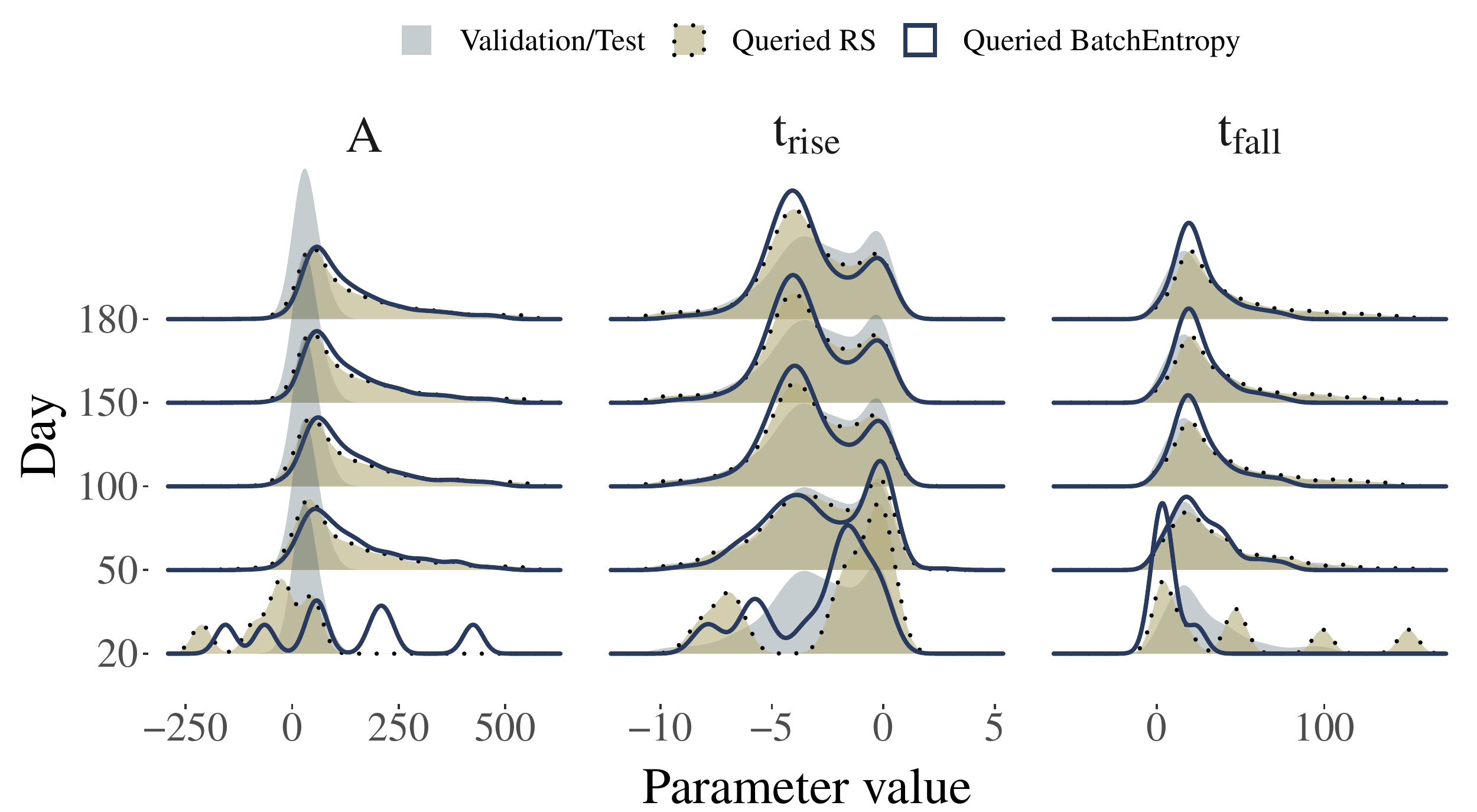}
    \caption{Distribution in $r$-band feature space for 3 different samples: Validation (filled grey), Random Sampling (dotted filled brown), and  BatchEntropy (solid dark blue line) as a function of days since the start of the survey (20 to 180, from bottom to top). This results correspond to an initial training of 10 objects. 
    }
  \label{fig:joyQd}
\end{figure*}

In order to test if this behavior is derived from the biases known to exist in the original training, we applied the same learning strategies to the case where the initial training sample is composed of only 10 objects randomly chosen from the original SNPCC spectroscopic sample (5 SNe Ia and 5 non-Ia). 
The evolution of all metrics is shown in Figure \ref{fig:metric} (right  column) and numerical values for their final state are given in Table~\ref{tab:train10}. In this scenario, the initial classifier does not contain much information; accuracy, purity and FoM start with lower values. Nevertheless, they quickly improve with each iteration, achieving results as good as those obtained in the previous case. At the final stage, the training samples contain $\approx$ 1,810 objects. Since a small initial training is less biased and more sensitive to the addition of new data, the active learning strategies clearly outperforms RS. The best performing active learning strategy (BatchEntropy) achieved a FoM of $0.34$, while RS delivered a FoM of $0.27$, a difference of $\approx 26\%$ ($0.07$) and an increase of 75\% when compared to the difference between USE and RS in the previous case case ($0.04$). This increase comes from a 28\% increase in efficiency delivered by BatchEntropy over RS. Figure~\ref{fig:joyQd} shows the evolution in feature space of the samples queried by RS and BatchEntropy in comparison to the validation/test samples. Comparing Figures \ref{fig:joyQ} and \ref{fig:joyQd} it is clear that both strategies (RS and BatchEntropy) evolve the queried sample towards the validation set but subjected to the constraints of the available pool sample at each iteration. 

%===============================================================================================
\section{Summary and Conclusions}
\label{sec:conclusions}

Active learning strategies are  promising techniques used to construct optimal training samples given scarce labeling resources. Nevertheless, stress tests probing their robustness under realistic conditions are largely missing in the literature.  In many real world situations, the assumptions of sample representativeness or stability between samples are hard to meet, though the necessity to optimize the allocation of labeling  resources is paramount.  

In this work we focus on the classification of a subclass of extragalactic astronomical transients:  supernovae. While this issue has received great attention in the last decade \cite{kessler2010, Ishida2013, lochner2016, ishida2019b, moller2020}, the community is still far from developing a completely automated system able to optimize the allocation of spectroscopic follow-up resources. In this work, we build upon the efforts reported in \cite{Ishida2019} and present for the first time a simulated data environment which simultaneously takes into account: 1) the necessity to estimate the current brightness of an object in order to make a decision about spectroscopic follow-up (only partial information is available at the time of labeling), 2) the evolution of training and pool samples with time, 3) the spectroscopic time required to observe each object in 2 different telescopes as a function of time (different labeling costs per day, object and budget source) and 4) the limited telescope time available per night (knapsack constraints).

We tested the performance of random sampling (RS) as well as three batch active learning strategies based on uncertainty sampling. When using the original training sample provided within the SNPCC data set (1,103 objects) as a starting point, active learning strategies did not significantly improve upon RS. This is a direct consequence of the biases known to exist between spectroscopic and photometric samples, combined with the large size of the initial training set, and the limited number of available nights (active learning iterations). Given these constraints, we constructed a second data scenario with a very small initial training set (10 objects). This initial state contained a negligible amount of information, but it was unbiased and highly sensitive to additional samples. Here, all active learning strategies clearly outperformed RS results. The best strategy (BatchEntropy) improved by 26\% the results delivered by RS. The small initial training set achieved the same figure of merit using 1,093 fewer spectroscopically confirmed light curves (labels).

Such results emphasize the importance of planning, in advance, the construction of training samples for machine learning applications. By delegating the complete construction of the training sample to the active learning algorithm, we can ensure optimal classification results and obviate the use of legacy data or the need to  model discrepancies between traditional spectroscopic and photometric samples. 

Moreover, we showed that active learning strategies are robust in the presence of complex and realistic constraints on data collection. However, the fact that different batch strategies presented similar behavior indicates that our current techniques for acquiring diverse committees can be improved. This is an important issue which will be addressed in a future work. 

Finally, we recognize that the scenario presented in this work is still incomplete. We failed to take into account: uncertainties due to our feature extraction method and the extrapolated brightness used to calculate the cost of each observation\footnote{This can be generalized as uncertainty in our data points, which has been studied mainly from a theoretical perspective with very ideal types of classifiers and noise models \cite{nowak2009noisy, nowak2011geometry}.}; the probability that a labeling request is not fulfilled or that it may be incorrect; the impact of the resulting classifications in further scientific results and observational effects like airmass (position of a given source in the sky) and weather conditions (e.g. seeing, cloud cover). This complex environment makes the classification of transient astronomical sources an excellent test bench for developing learning algorithms. These are all crucial issues which will shape the scientific results from the next generation of large scale astronomical surveys and, consequently, our understanding of the Universe.

%===============================================================================================
\section*{Acknowledgment}

%Put sponsor acknowledgments in the unnumbered footnote on the first page.

The authors would like to thank David Kirkby and Connor Sheere for insightful discussions. This work is part of the Recommendation System for Spectroscopic Follow-up (RESSPECT) project, governed by an inter-collaboration agreement signed between the Cosmostatistics Initiative (COIN) and the LSST Dark Energy Science Collaboration (DESC). This research is supported in part by the HPI Research Center in Machine Learning and Data Science at UC Irvine. EEOI and SS acknowledge financial support from CNRS 2017 MOMENTUM grant under the project \textit{Active Learning for Large Scale Sky Surveys}.  SGG and AKM acknowledge support by FCT under Project CRISP PTDC/FIS-AST-31546/2017. This work was partly supported by the Hewlett Packard Enterprise Data Science Institute (HPE DSI) at the University of Houston. DOJ is supported by a Gordon and Betty Moore Foundation postdoctoral fellowship at the University of California, Santa Cruz.  Support for this work was provided by NASA through the NASA Hubble Fellowship grant HF2-51462.001 awarded by the Space Telescope Science Institute, which is operated by the Association of Universities for Research in Astronomy, Inc., for NASA, under contract NAS5-26555. BQ is supported by the International Gemini Observatory, a program of NSF’s NOIRLab, which is managed by the Association of Universities for Research in Astronomy (AURA) under a cooperative agreement with the National Science Foundation, on behalf of the Gemini partnership of Argentina, Brazil, Canada, Chile, the Republic of Korea, and the United States of America. AIM acknowledges support from the Max Planck Society and the Alexander von Humboldt Foundation in the framework of the Max Planck-Humboldt Research Award endowed by the Federal Ministry of Education and Research. L.G. was funded by the European Union's Horizon 2020 research and innovation programme under the Marie Sk\l{}odowska-Curie grant agreement No. 839090. This work has been partially supported by the Spanish grant PGC2018-095317-B-C21 within the European Funds for Regional Development (FEDER).

\appendix
\subsection{BALD equivalence to Average KL-divergence}
\label{ap:BALD}

Here we show the equivalence between the BALD objective and the average KL-Divergence. We start with the BALD objective, which is the mutual information between the model's parameters and the target label of a given data point. 

\renewcommand{\qedsymbol}{\rule{0.5em}{0.5em}}

%\begin{proof}
%\begin{equation*}
%\begin{split}
\begin{eqnarray}
    I(\theta, y | x, \mathcal{D}) & = & H(p(y | x, \mathcal{D})) - \mathbb{E}_{p(\theta | \mathbf{D})}[H(p(y | x, \theta))] \nonumber \\
    & = & H(\mathbb{E}_{p(\theta | \mathbf{D})}[H(p(y | x, \theta))]) - \nonumber \\
     & & \mathbb{E}_{p(\theta | \mathbf{D})}[H(p(y | x, \theta))] \nonumber \\
    & \approx &  H ( \dfrac{1}{C} \sum_{c} P_{\theta_c} (y | x) ) -  \nonumber \\
    & & . \qquad \dfrac{1}{C} \sum_{c} H( p(y | x, \theta_{c})) \nonumber \\
    & = & \dfrac{1}{C} \sum_{c} \sum_{y}  P_{\theta_c} (y | x) \log\left(\dfrac{P_{\theta_c} (y | x)}{P_{C} (y | x)}\right) \nonumber \\
    & = & \dfrac{1}{C} \sum_{c} KL(P_{\theta_{c}}(y|x) ||  P_{C}(y|x))
\end{eqnarray}
%\end{split}
%\end{equation*}
%\end{proof}

Where the approximate equality is because we can only take finitely many samples from the posterior distribution of the model parameters. Hence we have shown when one can only compute Monte Carlo estimates of the BALD objective it is equivalent the average KL-divergence objective.

%===============================================================================================

% Equations 

%\begin{equation}
%    \sum_{c} p(y_{c}|x)\log p(y_{c}|x)
%\end{equation}

\bibliographystyle{IEEEtran}
\bibliography{ref.bib}

\begin{thebibliography}{10}
\providecommand{\url}[1]{#1}
\csname url@samestyle\endcsname
\providecommand{\newblock}{\relax}
\providecommand{\bibinfo}[2]{#2}
\providecommand{\BIBentrySTDinterwordspacing}{\spaceskip=0pt\relax}
\providecommand{\BIBentryALTinterwordstretchfactor}{4}
\providecommand{\BIBentryALTinterwordspacing}{\spaceskip=\fontdimen2\font plus
\BIBentryALTinterwordstretchfactor\fontdimen3\font minus
  \fontdimen4\font\relax}
\providecommand{\BIBforeignlanguage}[2]{{%
\expandafter\ifx\csname l@#1\endcsname\relax
\typeout{** WARNING: IEEEtran.bst: No hyphenation pattern has been}%
\typeout{** loaded for the language `#1'. Using the pattern for}%
\typeout{** the default language instead.}%
\else
\language=\csname l@#1\endcsname
\fi
#2}}
\providecommand{\BIBdecl}{\relax}
\BIBdecl

\bibitem{Settles12}
B.~Settles, \emph{Active Learning}.\hskip 1em plus 0.5em minus 0.4em\relax
  Morgan \& Claypool, 2012.

\bibitem{Ivezic2014}
Z.~Ivezi\'c, A.~J. Connolly, J.~T. VanderPlas, and A.~Gray, \emph{Statistics,
  Data Mining, and Machine Learning in Astronomy: A Practical Python Guide for
  the Analysis of Survey Data}.\hskip 1em plus 0.5em minus 0.4em\relax
  Princeton University Press, 2014.

\bibitem{Beck2017}
\BIBentryALTinterwordspacing
R.~{Beck}, C.~A. {Lin}, E.~E.~O. {Ishida}, F.~{Gieseke}, R.~S. {de Souza},
  M.~V. {Costa-Duarte}, M.~W. {Hattab}, and A.~{Krone-Martins}, ``{On the
  realistic validation of photometric redshifts},'' \emph{Monthly Notices of
  the Royal Astronomical Society}, vol. 468, no.~4, pp. 4323--4339, 03 2017.
  [Online]. Available: \url{https://doi.org/10.1093/mnras/stx687}
\BIBentrySTDinterwordspacing

\bibitem{Ishida2019}
E.~E.~O. {Ishida}, R.~{Beck}, S.~{Gonz{\'a}lez-Gait{\'a}n}, R.~S. {de Souza},
  A.~{Krone-Martins}, J.~W. {Barrett}, N.~{Kennamer}, and e.~{Vilalta},
  ``{Optimizing spectroscopic follow-up strategies for supernova photometric
  classification with active learning},'' \emph{Monthly Notices of the Royal
  Astronomical Society}, vol. 483, no.~1, pp. 2--18, Feb. 2019.

\bibitem{solorio2005}
T.~{Solorio}, O.~{Fuentes}, R.~{Terlevich}, and E.~{Terlevich}, ``{An active
  instance-based machine learning method for stellar population studies},''
  \emph{Monthly Notices of the Royal Astronomical Society}, vol. 363, pp.
  543--554, Oct. 2005.

\bibitem{richards2012b}
J.~W. {Richards}, D.~L. {Starr}, H.~{Brink}, A.~A. {Miller}, J.~S. {Bloom},
  N.~R. {Butler}, J.~B. {James}, J.~P. {Long}, and {et al.}, ``{Active Learning
  to Overcome Sample Selection Bias: Application to Photometric Variable Star
  Classification},'' \emph{The Astrophysical Journal}, vol. 744, p. 192, Jan.
  2012.

\bibitem{riess1998}
A.~G. {Riess}, A.~V. {Filippenko}, P.~{Challis}, A.~{Clocchiatti},
  A.~{Diercks}, P.~M. {Garnavich}, R.~L. {Gilliland}, C.~J. {Hogan}, and {et
  al.}, ``{Observational Evidence from Supernovae for an Accelerating Universe
  and a Cosmological Constant},'' \emph{The Astronomical Journal}, vol. 116,
  pp. 1009--1038, Sep. 1998.

\bibitem{perlmutter1999}
S.~{Perlmutter}, G.~{Aldering}, G.~{Goldhaber}, R.~A. {Knop}, P.~{Nugent},
  P.~G. {Castro}, S.~{Deustua}, S.~{Fabbro}, A.~{Goobar}, and {et al.},
  ``{Measurements of {$\Omega$} and {$\Lambda$} from 42 High-Redshift
  Supernovae},'' \emph{The Astrophysical Journal}, vol. 517, pp. 565--586, Jun.
  1999.

\bibitem{Phillips1987}
M.~M. {Phillips}, A.~C. {Phillips}, S.~R. {Heathcote}, V.~M. {Blanco},
  D.~{Geisler}, D.~{Hamilton}, N.~B. {Suntzeff}, F.~J. {Jablonski}, and {et
  al.}, ``{The type IA supernova 1986G in NGC 5128 : optical photometry and
  spectra.}'' \emph{Publications of the Astronomical Society of the Pacific},
  vol.~99, pp. 592--605, Jul. 1987.

\bibitem{foster2016}
F.~{F{\"o}rster}, J.~C. {Maureira}, J.~{San Mart{\'\i}n}, M.~{Hamuy},
  J.~{Mart{\'\i}nez}, P.~{Huijse}, G.~{Cabrera}, L.~{Galbany}, and {et al.},
  ``{The High Cadence Transient Survey (HITS). I. Survey Design and Supernova
  Shock Breakout Constraints},'' \emph{The Astrophysical Journal}, vol. 832,
  no.~2, p. 155, Dec. 2016.

\bibitem{kessler2010}
R.~{Kessler}, B.~{Bassett}, P.~{Belov}, V.~{Bhatnagar}, H.~{Campbell},
  A.~{Conley}, J.~A. {Frieman}, A.~{Glazov}, and {et al.}, ``{Results from the
  Supernova Photometric Classification Challenge},'' \emph{Publications of the
  Astronomical Society of the Pacific}, vol. 122, no. 898, p. 1415, Dec. 2010.

\bibitem{snana}
R.~{Kessler}, J.~P. {Bernstein}, D.~{Cinabro}, B.~{Dilday}, J.~A. {Frieman},
  S.~{Jha}, S.~{Kuhlmann}, G.~{Miknaitis}, M.~{Sako}, M.~{Taylor}, and
  J.~{Vanderplas}, ``{SNANA: A Public Software Package for Supernova
  Analysis},'' \emph{Publications of the Astronomical Society of the Pacific},
  vol. 121, no. 883, p. 1028, Sep. 2009.

\bibitem{bazin2009}
G.~{Bazin}, N.~{Palanque-Delabrouille}, J.~{Rich}, V.~{Ruhlmann-Kleider},
  E.~{Aubourg}, L.~{Le Guillou}, P.~{Astier}, C.~{Balland}, and {et al.},
  ``{The core-collapse rate from the Supernova Legacy Survey},''
  \emph{Astronomy and Astrophysics}, vol. 499, no.~3, pp. 653--660, Jun. 2009.

\bibitem{Breiman01}
L.~Breiman, ``Random forests,'' \emph{Machine Learning}, vol.~45, no.~1, pp. 5
  -- 32, 2001.

\bibitem{krause2008near}
A.~Krause, A.~Singh, and C.~Guestrin, ``Near-optimal sensor placements in
  gaussian processes: Theory, efficient algorithms and empirical studies,''
  \emph{Journal of Machine Learning Research}, vol.~9, no. Feb, pp. 235--284,
  2008.

\bibitem{krause2014submodular}
A.~Krause and D.~Golovin, ``Submodular function maximization.''

\bibitem{freund1997selective}
Y.~Freund, H.~S. Seung, E.~Shamir, and N.~Tishby, ``Selective sampling using
  the query by committee algorithm,'' \emph{Machine Learning}, vol.~28, no.
  2-3, pp. 133--168, 1997.

\bibitem{seung1992statistical}
H.~S. Seung, H.~Sompolinsky, and N.~Tishby, ``Statistical mechanics of learning
  from examples,'' \emph{Physical review A}, vol.~45, no.~8, p. 6056, 1992.

\bibitem{breiman1996bagging}
L.~Breiman, ``Bagging predictors,'' \emph{Machine learning}, vol.~24, no.~2,
  pp. 123--140, 1996.

\bibitem{mackay2003information}
\BIBentryALTinterwordspacing
D.~MacKay, D.~Kay, and C.~U. Press, \emph{Information Theory, Inference and
  Learning Algorithms}.\hskip 1em plus 0.5em minus 0.4em\relax Cambridge
  University Press, 2003. [Online]. Available:
  \url{https://books.google.fr/books?id=AKuMj4PN\_EMC}
\BIBentrySTDinterwordspacing

\bibitem{kullback1951information}
S.~Kullback and R.~A. Leibler, ``On information and sufficiency,'' \emph{The
  Annals of Mathematical Statistics}, vol.~22, no.~1, pp. 79--86, 1951.

\bibitem{houlsby2011bayesian}
N.~Houlsby, F.~Husz{\'a}r, Z.~Ghahramani, and M.~Lengyel, ``Bayesian active
  learning for classification and preference learning,'' \emph{arXiv preprint
  arXiv:1112.5745}, 2011.

\bibitem{nemhauser1978analysis}
G.~L. Nemhauser, L.~A. Wolsey, and M.~L. Fisher, ``An analysis of
  approximations for maximizing submodular set functions—i,''
  \emph{Mathematical programming}, vol.~14, no.~1, pp. 265--294, 1978.

\bibitem{kirsch2019batchbald}
A.~Kirsch, J.~van Amersfoort, and Y.~Gal, ``Batchbald: Efficient and diverse
  batch acquisition for deep bayesian active learning,'' in \emph{Advances in
  Neural Information Processing Systems}, 2019, pp. 7026--7037.

\bibitem{yeung1991new}
R.~W. Yeung, ``A new outlook on shannon's information measures,'' \emph{IEEE
  transactions on information theory}, vol.~37, no.~3, pp. 466--474, 1991.

\bibitem{krause2008optimizing}
A.~Krause, ``Optimizing sensing,'' Ph.D. dissertation, Carnegie Mellon
  University, 2008.

\bibitem{Ishida2013}
\BIBentryALTinterwordspacing
E.~E.~O. Ishida and R.~S. de~Souza, ``{Kernel PCA for Type Ia supernovae
  photometric classification},'' \emph{Monthly Notices of the Royal
  Astronomical Society}, vol. 430, no.~1, pp. 509--532, 01 2013. [Online].
  Available: \url{https://doi.org/10.1093/mnras/sts650}
\BIBentrySTDinterwordspacing

\bibitem{lochner2016}
M.~{Lochner}, J.~D. {McEwen}, H.~V. {Peiris}, O.~{Lahav}, and M.~K. {Winter},
  ``{Photometric Supernova Classification with Machine Learning},'' \emph{The
  Astrophysical Journal Supplements}, vol. 225, p.~31, Aug. 2016.

\bibitem{ishida2019b}
E.~E.~O. {Ishida}, ``{Machine learning and the future of supernova
  cosmology},'' \emph{Nature Astronomy}, vol.~3, pp. 680--682, Aug. 2019.

\bibitem{moller2020}
A.~{M{\"o}ller} and T.~{de Boissi{\`e}re}, ``{SuperNNova: an open-source
  framework for Bayesian, neural network-based supernova classification},''
  \emph{Monthly Notices of the Royal Astronomical Society}, vol. 491, no.~3,
  pp. 4277--4293, Jan. 2020.

\bibitem{nowak2009noisy}
R.~Nowak, ``Noisy generalized binary search,'' in \emph{Advances in Neural
  Information Processing Systems}, 2009, pp. 1366--1374.

\bibitem{nowak2011geometry}
R.~D. Nowak, ``The geometry of generalized binary search,'' \emph{IEEE
  Transactions on Information Theory}, vol.~57, no.~12, pp. 7893--7906, 2011.

\end{thebibliography}

\end{document}